# Coherent motion of stereocilia assures the concerted gating of hair-cell transduction channels

Andrei S. Kozlov[1,2], Thomas Risler[1,2,3] & A. J. Hudspeth[1]

**The hair cell's mechanoreceptive organelle, the hair bundle, is highly sensitive because its transduction channels open over a very narrow range of displacements. The synchronous gating of transduction channels also underlies the active hair-bundle motility that amplifies and tunes responsiveness. The extent to which the gating of independent transduction channels is coordinated depends on how tightly individual stereocilia are constrained to move as a unit. Using dual-beam interferometry in the bullfrog's sacculus, we found that thermal movements of stereocilia located as far apart as a bundle's opposite edges display high coherence and negligible phase lag. Because the mechanical degrees of freedom of stereocilia are strongly constrained, a force applied anywhere in the hair bundle deflects the structure as a unit. This feature assures the concerted gating of transduction channels that maximizes the sensitivity of mechanoelectrical transduction and enhances the hair bundle's capacity to amplify its inputs.**

[1]Howard Hughes Medical Institute and Laboratory of Sensory Neuroscience, The Rockefeller University, 1230 York Avenue, New York, New York 10021, USA. [2]These authors contributed equally to this work. [3]Present address: Laboratoire Physicochimie Curie UMR 168, Institut Curie section de Recherche, 26 rue d'Ulm, F-75248 Paris cedex 05, France. Correspondence should be addressed to A.J.H. (hudspaj@rockefeller.edu).

The high sensitivity of sensory systems requires an efficient use of the energy in stimuli to bias the open probability of ion channels. For a hair cell of the inner ear, mechanical forces directly gate transduction channels atop stereocilia, the rod-like constituents of the mechanosensitive hair bundle[1]. A hair bundle's sensitivity is determined by the relation between the applied force and the number of channels opened: the narrower the force range over which gating occurs, the greater the sensitivity. The coordinated gating of transduction channels is also thought to underlie active hair-bundle motility, a component of the active process that amplifies and tunes the responses of hair cells[2].

Because mechanical stimuli are ordinarily applied at the tall edge of a hair bundle, channel gating depends upon the propagation of mechanical force across the array of stereocilia. Each stereocilium possesses a basal rootlet of actin filaments that tends to hold the process upright; as measured at the hair bundle's tip, the combined stiffness of these stereociliary pivots is about 200 $\mu N \cdot m^{-1}$ (ref.[3]). In addition, the successive stereocilia in each file are joined by tip links that are thought to represent the gating springs attached to transduction channels at one or both ends. For large bundle deflections and in the presence of a physiological concentration of $Ca^{2+}$, the combined stiffness of these gating springs is typically 1000 $\mu N \cdot m^{-1}$ (ref. 3). The opening and closing of transduction channels reduces the effective stiffness of the gating springs—the phenomenon of gating compliance—to a value comparable to that of the pivots, or even lower[4,5].

The stereocilia of a hair bundle appear at first glance to be connected in a series-parallel configuration such that a force applied to the tallest stereocilium in each file would first deflect that process alone (**Fig. 1a**). Movement of the tallest stereocilium would then tighten the tip link and perhaps other filaments connecting it to the second, deflecting that process; the second stereocilium would in turn pull on the third, and so on across the hair bundle. If stereocilia were to operate in such a configuration, the magnitude of their deflection would diminish progressively from the tallest to the shortest[6,7] and the open probability of the transduction channels would therefore vary with their location (**Fig. 1b**). This arrangement would broaden the relation between force and channel open probability and thereby reduce the sensitivity of mechanoelectrical transduction. The ensemble of mechanosensitive channels would function more efficiently if the stereocilia were instead to adopt a parallel arrangement, in which an applied force is shared equally among them and deflects all to a similar extent (**Fig. 1b**). This arrangement would prevail if the stereocilia were somehow constrained to remain in close contact during excitation.

Visual observations[8] and video measurements[9,10] have suggested that a hair bundle moves as a unit when subjected to relatively large, low-frequency stimuli. However, a conclusive examination of the propagation of mechanical forces across a bundle requires simultaneous measurements of the positions of two stereocilia with a sub-nanometer spatial and a sub-millisecond temporal resolution, for these are the scales typical of stereociliary movements during hearing. Because interferometry can detect motions with the requisite precision, we constructed a dual-beam, differential laser interferometer and used it to examine the correlations between the thermal motions of individual stereocilia both in quiescent and in spontaneously oscillating hair bundles (**Fig. 1c**). Our experiments showed that the movements of stereocilia located as far apart as the bundle's opposite edges displayed high coherence and zero phase lag over a wide range of frequencies. This result implies that stereocilia are strongly constrained to move in unison, a feature that promotes the high sensitivity of mechanoelectrical transduction and fosters the coordinated gating that underlies active hair-bundle motility.

## RESULTS

The morphological organization of the hair bundle is highly conserved throughout the vertebrates. The experiments reported in this study were therefore performed on hair cells from the bullfrog's sacculus, a preparation whose mechanical and electrical properties have been extensively studied and which was used successfully in earlier experiments employing laser interferometery[11]. For each hair bundle, we measured the time-dependent positions of stereocilia with three distinct configurations of the dual-beam interferometer.

In the first arrangement, the two laser beams were focused together on one of the bundle's edges, for example at the same point on the tip of the longest stereocilium (**Fig. 2a**). As expected, the resultant records were closely similar to one another (**Fig. 2b**). For a quiescent hair bundle, the autocorrelation of each signal declined exponentially from its peak value with a time constant of approximately 1 ms (**Fig. 2c**). This correlation time represents the quotient of the bundle's drag coefficient, *circa* 130 nN·s·m$^{-1}$ (ref. 4,11), and its stiffness, about 150 µN·m$^{-1}$ including the gating compliance near the resting position[12]. In 38 measurements from 18 hair bundles, the peak value of the cross-correlation of the two signals was $0.93 \pm 0.04$. The difference of 7% between this value and the maximal value of one provides a quantitative indication of the noise level in the illumination and detection systems.

A second experimental configuration provided a control for the presence of drift by the whole epithelium, which might have introduced artifactual correlations between the motions of two different spots on a given hair bundle. In this instance, we acquired 17 measurements with the laser beams focused on two adjacent hair bundles. As expected for signals from independently moving sources, the cross-correlation was randomly distributed in time and displayed an average magnitude of $0.08 \pm 0.03$, similar to that of the background noise (**Fig. 2c**).

In the third and critical set of measurements, we estimated the degree of common motion between stereocilia across a hair bundle by positioning the laser beams on its two opposite edges. The individual records were highly similar in quiescent hair bundles (**Fig. 3a**). The autocorrelations and cross-correlation from each cell were almost identical (**Fig. 3b**). For 29 measurements from 18 quiescent hair bundles, the peak value of the averaged cross-correlation was $0.92 \pm 0.03$; this value differed by only 1% from the peak cross-correlation obtained with both lasers focused on the same stereocilium.

The cross-correlation commingles information about the components of hair-bundle movement over the full measurable range of frequencies. To assess the degree of common motion as a function of frequency, we characterized the coherency spectrum of the two signals. The coherency is a dimensionless, complex quantity whose argument represents the estimated phase lag between two signals and whose modulus, called the coherence, represents the quality of this phase estimate[13]. For signals of infinite duration, the coherence is bounded at each frequency by zero for uncorrelated signals and by unity for perfectly correlated ones. For 18 quiescent hair bundles, and except for a few deviations caused by laser noise, the average coherence obtained in 29 measurements from opposite hair-bundle edges remained high at all sampled frequencies and exceeded 0.88 between 100 Hz and 5 kHz (**Fig. 3c**). When the two laser beams were positioned together on one edge of the bundle, the average coherence in 38 measurements was only 2% greater (**Fig. 3d**). Phase-lag estimates were close to zero over the whole spectrum for all the quiescent hair bundles. Between 100 Hz and 5 kHz, the standard deviation was smaller than 0.23 radian for determinations from the bundles' opposite edges and beneath 0.18 radian for measurements at the same edges. The coherence and phase estimates were limited at low frequencies by independent drifts in the two laser beams and at high frequencies by the diminishing signal power, which eventually approached the background noise level.

The gating compliance of healthy hair bundles lowers their effective stiffness to a value comparable to that of a damaged bundle. To ensure that our data reflected the activity of functional hair bundles, we repeated the measurements with nine oscillatory bundles, whose spontaneous movements required the normal gating of transduction channels by intact tip links (**Fig. 4a**). In this instance, the average cross-correlation obtained in 29 measurements from opposite hair-bundle edges peaked at a value of $0.92 \pm 0.05$ (**Fig. 4b**). The associated coherence exceeded 0.87 between 100 Hz and 5 kHz, (**Fig. 4c**), as compared to the value of 0.93 obtained in 17 measurements with the lasers focused on the same bundle edge (**Fig. 4d**). In every instance, the estimated phase lag was negligible over the whole spectrum: the standard deviation was less than 0.25 radian for measurements at the opposite edges of the oscillating hair bundles, and below 0.12 radian for same-edge measurements.

The striking concordance in the movements of the stereocilia across a hair bundle might reflect the transmission of forces through the tip links or other filamentous connections between stereocilia. Alternatively, the geometrical arrangement of the hair bundle might constrain the stereocilia to move together. To distinguish between these possibilities, we treated cells with 5 mM BAPTA, a procedure known to disrupt tip links and certain other connections[14,15]. As expected because tip links contribute significantly to hair-bundle stiffness, treated bundles displayed both a greater root-mean-square displacement ($10.2 \pm 2.1$ nm, $n = 20$ $versus$ $5.5 \pm 1.8$ nm, $n = 67$; **Fig. 5a**) and an increased correlation time ($2.2 \pm 0.4$ ms, $n = 20$ $versus$ $1.1 \pm 0.4$ ms, $n = 67$; **Fig. 5b**). The coherence between 100 Hz and 5 kHz nonetheless exceeded 0.89 in 11 measurements from the opposite edges of the bundles (**Fig. 5c**) and 0.96 in nine measurements from the same edges (**Fig. 5d**). Phase-lag estimates showed standard deviations smaller than 0.14 radians for the opposite bundle edges and 0.08 radians for the same edges. That BAPTA neither lowered the coherence nor increased the phase lag implies that the coordination of motion among stereocilia does not depend on tip links.

To compare our findings with those anticipated for a series-parallel configuration of the hair bundle, we modeled the thermal motions expected for stereocilia interconnected by gating springs of various stiffnesses. For the stiffness value estimated in previous experiments on hair bundles from the bullfrog's sacculus[4,12], the cross-correlation between the movements of stereocilia on opposite edges of the modeled bundle was near zero (**Fig. 6a**). The coherence of these movements approached the value associated with independent motion at all frequencies; consistent with this result, the phase was essentially random (**Fig. 6b**). Even when the gating-spring stiffness was increased more than fiftyfold in an effort to foster closer coupling between stereocilia, the cross-correlation failed to attain the values found in actual experiments (**Fig. 6a**). In this instance, the coherency spectrum displayed coordinated movements of stereocilia on a bundle's opposite edges only at low frequencies, but essentially independent motion for frequencies in excess of 1 kHz (**Fig. 6c**). A simple series-parallel model therefore seems incapable of reproducing the experimental results with values of gating-spring stiffness compatible with the hair bundle's measured stiffness.

## DISCUSSION

The present results demonstrate that the stereocilia throughout a hair bundle exhibit strongly correlated motions up to high frequencies. The coherences measured with the two laser beams focused on the same or on opposite edges of hair bundles agree within 2% for frequencies between 100 Hz and 5 kHz. Because quiescent hair bundles display a root-mean-square motion of about 3 nm, these coherence values imply that the average splaying between their successive stereocilia is no more than a few tens of picometers. It follows that stimulus forces exerted on a hair bundle are distributed almost equally among the gating springs. Unlike a simple series-parallel arrangement, the bundle's actual configuration ensures that most of the stimulus energy is delivered to the gating machinery of transduction channels. The results also validate previous estimates of tip-link stiffness and buttress the model of negative hair-bundle stiffness, both of which rely on the assumption that the tip links are arrayed in parallel[4,16]. Although the present experiments involved a receptor organ sensitive to relatively low frequencies, we have observed similar behaviors in preliminary experiments performed on auditory hair cells in the gecko's cochlea.

What mechanisms might account for the high correlation between the motions of stereocilia at a bundle's opposite extremes? Viscosity could contribute to the phenomenon, for the tendency of any stereocilium to separate from a neighbor during high-frequency stimulation would be opposed by the resultant reduction in the fluid pressure between them[17]. Next, the basal and lateral links that conjoin stereocilia[18] might resist the separation of these processes. It should be noted, however, that the enzymatic digestion used for the present experimental preparation largely removed these attachments[19]. The most intriguing possibility is that stereocilia do not separate because they are forced together, whether at rest or in motion, by the curvature of the cuticular plate into which they insert[18]. If the stereocilia were prestressed against one another, the deflection of any stereocilium during stimulation would allow each successive stereocilium to relax towards its position of mechanical equilibrium. This movement would compel the hair bundle to move as a unit without imposing the increased stiffness associated with stereociliary cross-linking.

The negative stiffness of a hair bundle is analogous to the negative resistance of a neuron during the rising phase of an action potential: both depend upon the collective action of ion channels that are globally coupled to one another, whether mechanically or through the membrane voltage, and therefore display concerted gating. Such channel interactions resemble those between protein molecules involved in cooperative phenomena. Molecular cooperativity, such as that in hemoglobin and allosterically regulated enzymes, ordinarily involves molecular contacts so intimate that changes in the configuration of one subunit or domain are physically communicated to its neighbors. In a hair bundle, however, the stimulus-induced opening and closing of individual transduction channels—situated micrometers apart on separate stereocilia—alters the balance of forces in the whole structure, in turn affecting the force experienced by the entire channel ensemble.

## METHODS

**Experimental preparation.**   Sacculi dissected from adult bullfrogs (*Rana catesbeiana*) following a protocol approved by the Institutional Animal Care and Use Committee were maintained in oxygenated saline solution comprising 120 mM NaCl, 2 mM KCl, 1 mM CaCl$_2$, 10 mM D-glucose, and 5 mM HEPES at pH 7.3. After a 30–60 min digestion at room temperature in 1 mg·ml$^{-1}$ collagenase (type XI, Sigma Chemical Co.), each sensory epithelium was separated from the underlying connective tissue[11] and the otolithic membrane was removed. The epithelium was then folded in its plane of mirror symmetry[19] so that hair bundles protruded radially from the creased edge and could be imaged in profile. The preparation was secured to the coverslip bottom of an experimental chamber by placing over it a golden, 100-mesh electron-microscopic grid.

**Interferometric recording.** The dual-beam laser interferometer incorporated the basic features of a differential interferometer[11,20]. Two independent illumination pathways were established, one with red light from a 1.8-mW, 633-nm helium-neon laser (117A, Spectra-Physics), the other with green light from a 10-mW, 532-nm diode-pumped solid-state laser (85 GCA 010, Melles Griot Inc.). Each pathway included a Faraday reflection suppressor, a telescopic beam expander, and polarization and steering optics. The two beams were combined and directed into the microscope's differential-interference-contrast optical system, which split each circularly polarized beam into two orthogonally polarized components separated by *circa* 100 nm. A 40X objective lens of numerical aperture 0.8 acted as a condenser to focus each beam in the specimen plane to a diffraction-limited spot with a full width at half-maximal intensity of about 300 nm.

After traversing the specimen, the light from the two colored beams was collected through a second, identical objective lens and separated with a dichroic mirror into two independent measurement pathways. Each beam impinged on a polarizing beam splitter, which resolved the elliptically polarized light into two components whose magnitudes were measured with independent photodiodes. The amplified outputs were passed through eight-pole Butterworth anti-aliasing filters with a low-pass corner frequency of 20 kHz, then sampled with 16-bit resolution at 10-μs intervals. For each color channel, the ratio of the difference between the signals of orthogonal polarization to their sum was directly related to the phase difference between the two closely spaced optical pathways, and thus to the position of the object intercepted by the beam.

To calibrate each beam of the instrument, the lens forming one end of the associated beam-steering telescope was displaced through a known distance with a piezoelectrical manipulator. This procedure deflected the relevant focal spot in the specimen plane by a fixed amount set by the system's optical parameters. Using video imaging, we measured this shift in the specimen plane and thus determined the relation between the voltage applied to the piezoelectrical manipulator and the ensuing movement. Finally, during an actual experiment, the steering lens was displaced through a known distance and the resultant signal from the photodiodes was measured, thus establishing the calibration factor relating specimen movement to photodiode output.

In the absence of a specimen, each beam displayed a spectrally flat instrument noise of $5 \cdot 10^{-6}$ nm$^2 \cdot$Hz$^{-1}$ over the frequency range from 100 Hz to 20 kHz, corresponding to a root-mean-square sensitivity of 0.3 nm. When directed at one edge of a hair bundle, the interferometer detected thermal motion consistent with previous measurements[11,20-22]: the power-spectral density was approximately 0.03 nm$^2 \cdot$Hz$^{-1}$ up to a cutoff frequency near 180 Hz, implying a root-mean-square displacement of 3 nm. When the red and green laser beams were focused at the same point on a hair bundle, there was no detectable cross-talk between the two measurement channels.

**Data analysis.** Each measurement consisted of a set of twenty, 100-ms-long records sampled at 10-μs intervals, between which the detection apparatus was calibrated independently for each laser beam. Discrete Fourier analysis of these records was performed after tapering of the data with a Bartlett window. Correlation functions were estimated from Fourier power-spectral amplitudes for which tapered records had previously been padded with 100-ms-long strings of zeros to avoid spurious correlations. Outliers were rejected on the basis of unstable root-mean-square amplitudes, abnormal cross-correlation amplitudes, or pronounced drift in the raw data.

For each set of data in a given experimental condition, power spectra and correlation functions were computed with the remaining independent estimates. The averaged autocorrelation and cross-correlation functions were obtained after normalization of each record by its root-mean-square value. The coherence and phase spectra associated with these measurements represented the modulus and phase of the coherency spectrum defined at a frequency $f$ by

$$\gamma(f) = \langle C_{XY}(f) \rangle / \sqrt{\langle S_X(f) \rangle \langle S_Y(f) \rangle}$$

If $X(t)$ and $Y(t)$ represent the signals from the two laser beams for an individual record, $C_{XY}(f)$ denotes the associated cross-spectrum estimate at the frequency $f$, $S_X(f)$ and $S_Y(f)$ specify the two power-spectral estimates at that frequency, and $\langle ... \rangle$ indicates averaging over several independent records.

**Modeling.** To compare our findings with those expected if the stereocilia were to operate in the series-parallel configuration depicted in **Fig. 1b**, we simulated the outcome of the principal experiments. The modeled hair bundle consisted of a file of seven stereocilia, each subjected to linear frictional forces from the surrounding solution that included a deterministic component with an effective friction coefficient 20 nN·s·m$^{-1}$ and a stochastic white noise representing the thermal motion at an ambient temperature of 295 K. Each stereocilium was connected to the cuticular plate by a flexional spring of stiffness 5 μN·m$^{-1}$ and to each of its immediate neighbors by a gating spring whose stiffness was adjusted to either 1,000 μN·m$^{-1}$ or 53,000 μN·m$^{-1}$. The former value corresponds to previous estimates of gating-spring stiffness in the bullfrog's saccular hair cells[4,12]; the latter value is tenfold that proposed in a series-parallel model[7]. To mimic the experimental protocol, we generated data in sets of twenty, 100-ms-long records sampled at 10-μs intervals. We then analyzed the results by the procedures described above and averaged the results of ten independent repetitions.

## ACKNOWLEDGMENTS

The authors thank A. J. Hinterwirth for assistance in constructing the interferometer, B. Fabella for programming the experimental software, and O. Ahmad, M.O. Magnasco, K. Purpura, and J. Victor for useful discussions. The members of our research group provided helpful comments on the manuscript. The research reported in this paper was funded by the U.S. National Institutes of Health. T.R. was supported by funding from the F.M. Kirby Foundation and from the U.S. National Institutes of Health. A.S.K. was supported by Howard Hughes Medical Institute, of which A.J.H. is an Investigator.

## AUTHOR CONTRIBUTIONS

A.J.H. conceived the experiments, A.S.K. performed them, and T.R. conducted the data analysis. A.S.K., T.R., and A.J.H. wrote the paper.

## COMPETING INTERESTS STATEMENT

The authors declare that they have no competing financial interests.

---

**Figure 1** The hair bundle and its possible modes of motion. (**a, left**) A scanning electron micrograph depicts a saccular hair bundle, which comprises about 60 cylindrical stereocilia and a single kinocilium at its tall edge. (**a, right**) A schematic diagram of a slice along the central file of a hair bundle illustrates the geometrical arrangement of the stereocilia. The bundle is about 8 $\mu$m in height and each stereocilium is roughly 500 nm in diameter; tip links of exaggerated thickness are depicted between the adjacent stereocilia. Although the stereocilia differ substantially in height, the extensions of the associated tip links are nearly identical over the physiological range of stimulation[19]. The spots at the upper edges of the resting hair bundle represent the diffraction-limited laser beams employed in interferometric measurements, here positioned on the bundle's opposite edges. The calibration bar corresponds to 5 $\mu$m. (**b, left**) Upon application of a stimulus to the bundle (arrow), the force would spread decrementally from the longest to the progressively shorter stereocilia if these were arranged in the series-parallel configuration suggested by the bundle's structure. (**b, right**) By contrast, a principally parallel configuration would distribute the stimulus force more-or-less equally among the stereocilia. (**c, left**) Under thermal motion, the stereocilia would display relatively independent positional fluctuations in the series-parallel arrangement. (**c, right**) In the more parallel configuration, the stereocilia would be constrained to move together.

**Figure 2** Experimental preparation and control measurements (**a**) The laterally protruding hair bundles are apparent in a micrograph of a folded sensory epithelium secured by an electron-microscopic grid. The hair bundle selected for measurements, designated by a dashed square, is depicted at a higher magnification on the right. Each calibration bar corresponds to 10 $\mu$m. (**b**) When the green and red laser beams are focused at the same point on the tips of the longest stereocilia, the two records of thermal motion are highly similar. (**c**) The red dashed line represents the averaged

autocorrelation for 19 records acquired with the red laser beam, whereas the continuous blue line corresponds to the cross-correlation computed from the green- and red-laser records. The cross-correlation peaks at $0.97 \pm 0.02$, which indicates an instrumentation-noise level 3% of the peak signal. The black trace, which represents the cross-correlation averaged over 19 records with the laser beams focused on two adjacent bundles, lies at the system's noise level.

**Figure 3** Coherency of stereociliary motion for quiescent hair bundles. (**a**) When the green and red beams are positioned on the opposite edges of a hair bundle, the two traces are very similar apart from different amplitudes owing to different stereociliary lengths. (**b**) For this hair bundle, the cross-correlation, averaged over 20 records (blue line) and superimposed on the corresponding autocorrelation (red dashed line), reaches a peak value of $0.95 \pm 0.01$. (**c**) The averaged coherency spectrum for 29 measurements from the opposite edges of 18 hair bundles shows uniformly high values and a negligible phase lag at frequencies up to 10 kHz. The standard deviations for these measurements indicate the degree of variability in the different coherence and phase estimates. The phase spectrum shown at a higher magnification demonstrate a systematic deviation of the mean phase lag from zero at frequencies close to analog filters' cutoff. This deviation occurs in all records and results from non-identical phase delays introduced by the anti-aliassing filters. (**d**) In 38 measurements from the same cells, the spectrum for the beams focused on the same edge of the hair bundle is highly similar to that in (**c**).

**Figure 4** Coherency of stereociliary motion for oscillating hair bundles. (**a**) Although the interferometric records from the opposite edges of a spontaneously active hair bundle document displacements substantially larger than those observed in a quiescent bundle, the two traces are again very similar. (**b**) Averaged over 20 records, the cross-

correlation peaks at 0.99 ± 0.01.  The time constant of the correlations' decay exceeds that observed in quiescent cells, for example in **Fig. 3b**, owing to the low-frequency oscillatory movements of the bundles.  The negative values of the correlations for times exceeding ±7 ms reflect the bundle's tendency to move in the opposite direction after dwelling at one extreme or the other.  As shown on a coarser time scale in the insert, the cross-correlation also illustrates the roughly 70-Hz periodicity associated with the oscillations.  (**c**) In 29 measurements from nine hair bundles, the averaged spectrum obtained with the beams directed at the bundles' opposite edges displays a high coherence and negligible phase lag.  (**d**) The corresponding spectrum for 17 control measurements from the same hair bundles is essentially identical to the foregoing.

**Figure 5**  Effect of severing tip links on hair-bundle motion.  (**a**) Measurements from the opposite edges of a quiescent hair bundle portray the enhanced thermal motion after destruction of the tip links and other interstereociliary filaments by BAPTA.  (**b**) The cross-correlations before and after the treatment document the altered correlation time of the bundle's motion.  Severing the connections between stereocilia softened the hair bundle, thereby increasing the inversely related time constant for mechanical relaxation.  The peak cross-correlation in 20 records was 0.97 ± 0.02, to be compared with 0.89 ± 0.03 over 19 records for the same bundle prior to treatment.  (**c**) The coherency spectrum averaged across 11 measurements taken from four cells treated with BAPTA confirms the strong correlation between the movements of the bundles' opposite edges.  (**d**) The control spectrum for measurements from the same edges of the hair bundles differs negligibly from that in (**c**).

**Figure 6**  Modeling  of  hair-bundle  movements.  (**a**) Simulations  of  the  thermal movements expected for stereocilia in a series-parallel hair-bundle configuration, with the two indicated values of gating-spring stiffness, yield cross-correlations much lower

than those observed experimentally. (**b**) The coherency spectrum obtained for a gating-spring stiffness of 1,000 $\mu$N·m$^{-1}$ shows a nearly flat coherence near 0.2 and an essentially random phase for nearly all frequencies. This behavior, which is characteristic of finite samples of independent random processes, implies negligible coupling between stereocilia. (**c**) Even for an unrealistically large value of the gating-spring stiffness, 53,000 $\mu$N·m$^{-1}$, the coherency spectrum for a series-parallel model displays results very different from those observed experimentally. The averaged coherence lies below 0.8 for all frequencies exceeding 500 Hz and reaches the value expected for independent random processes around 1 kHz; the phase becomes nearly random above the latter frequency.

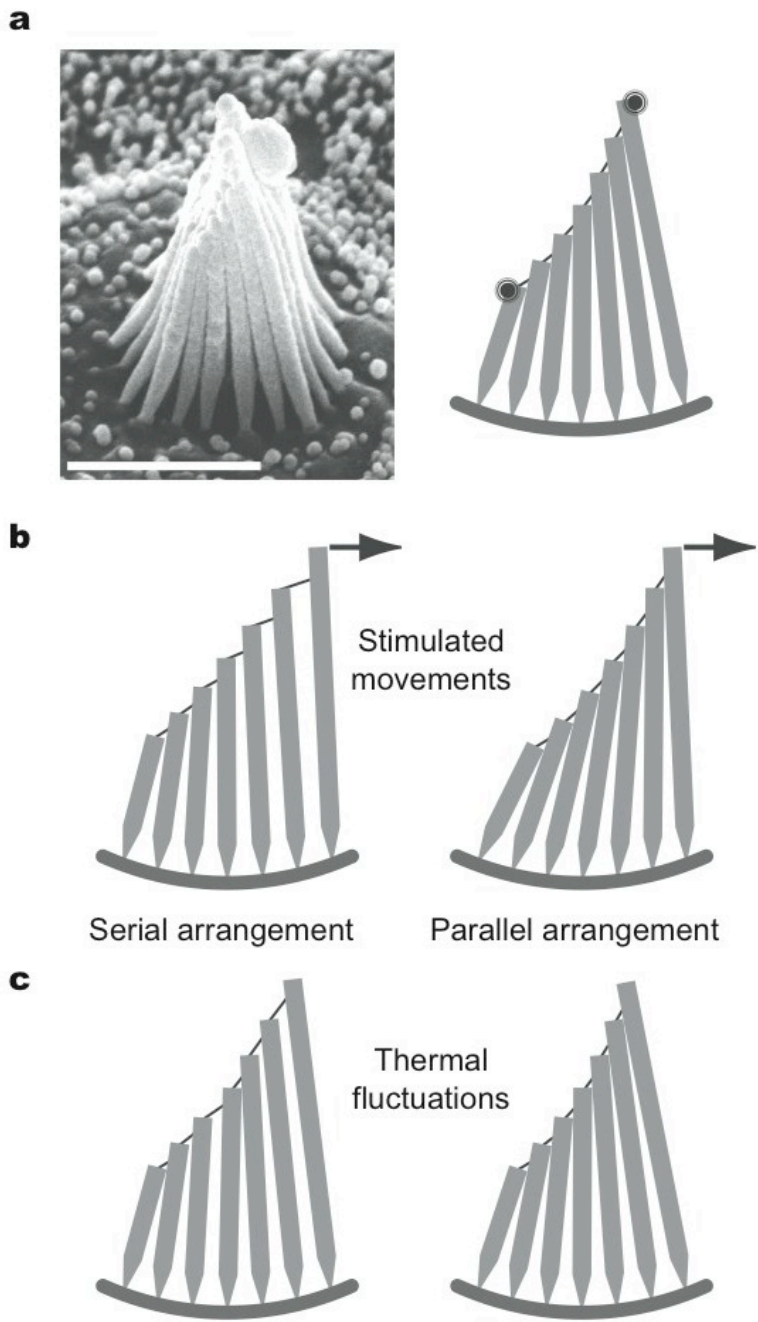

**a**

**b**

Stimulated movements

Serial arrangement    Parallel arrangement

**c**

Thermal fluctuations

Figure 1

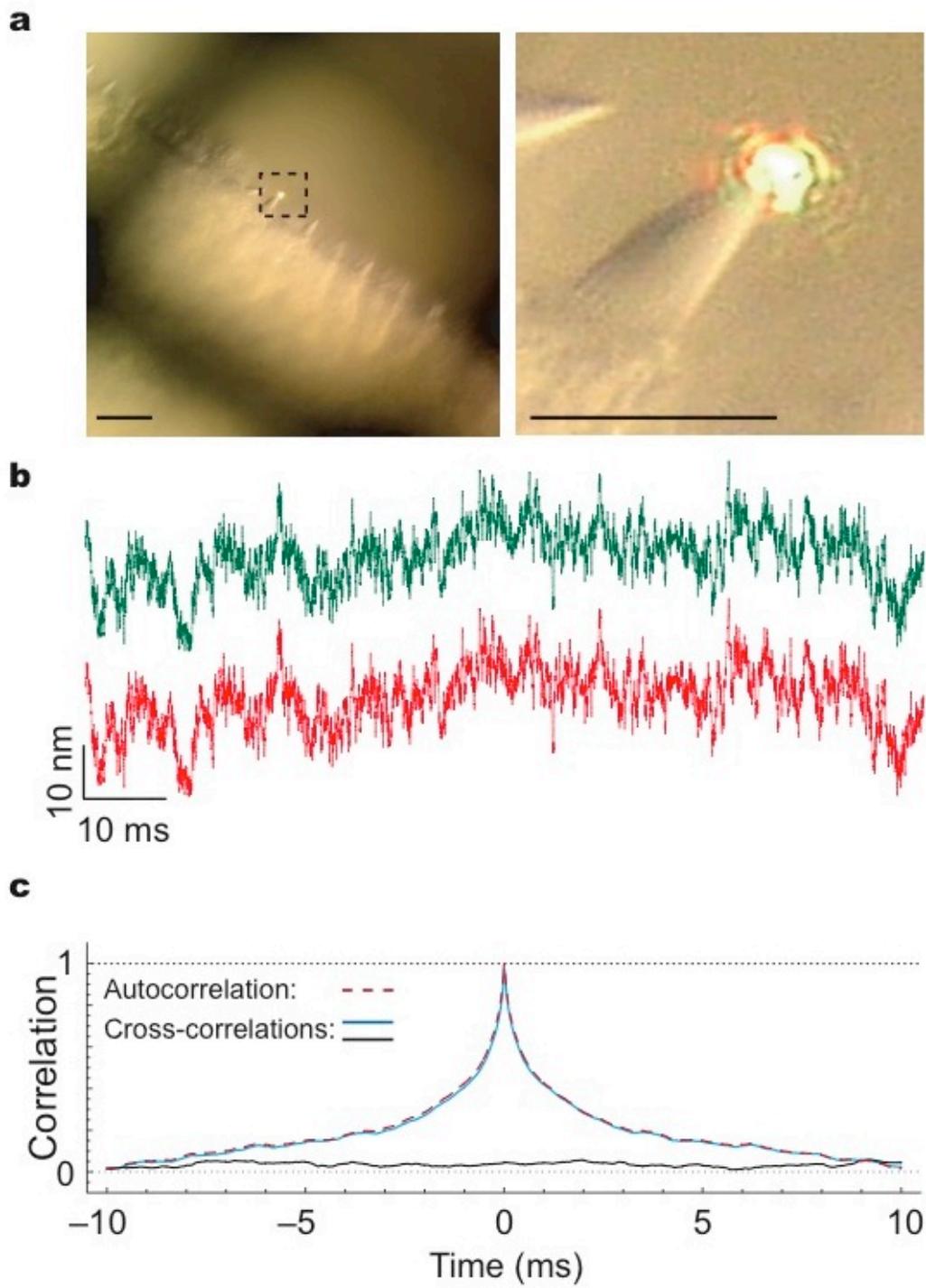

Figure 2

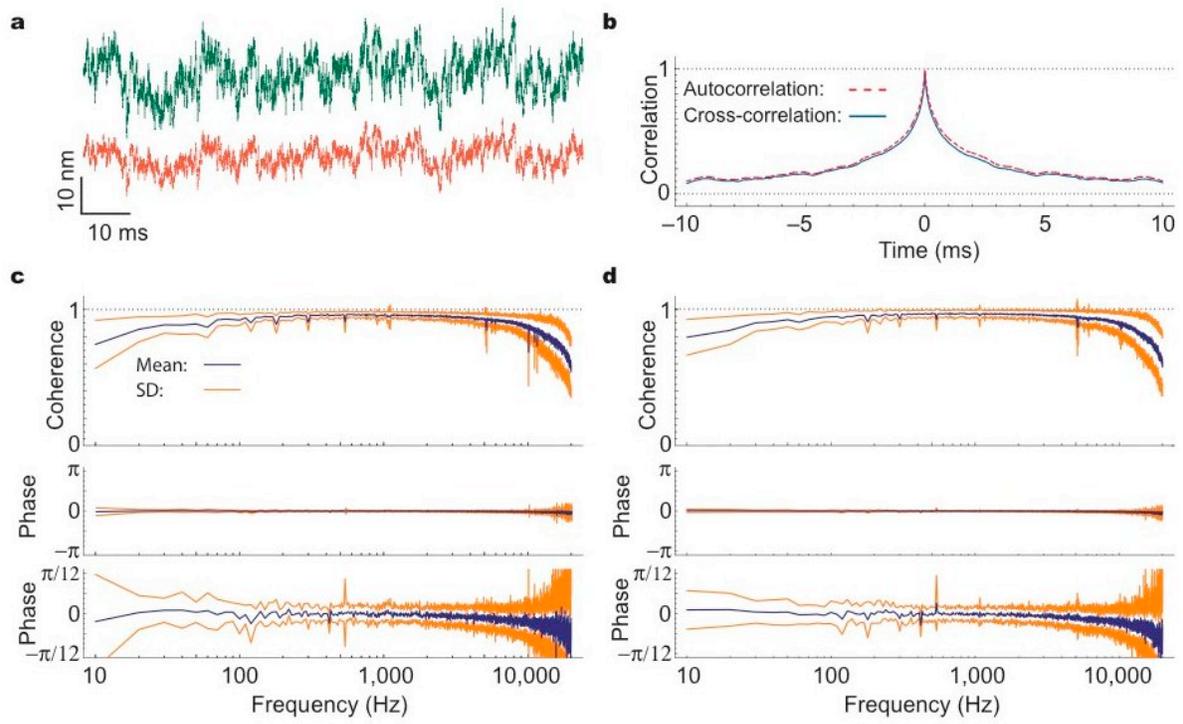

Figure 3

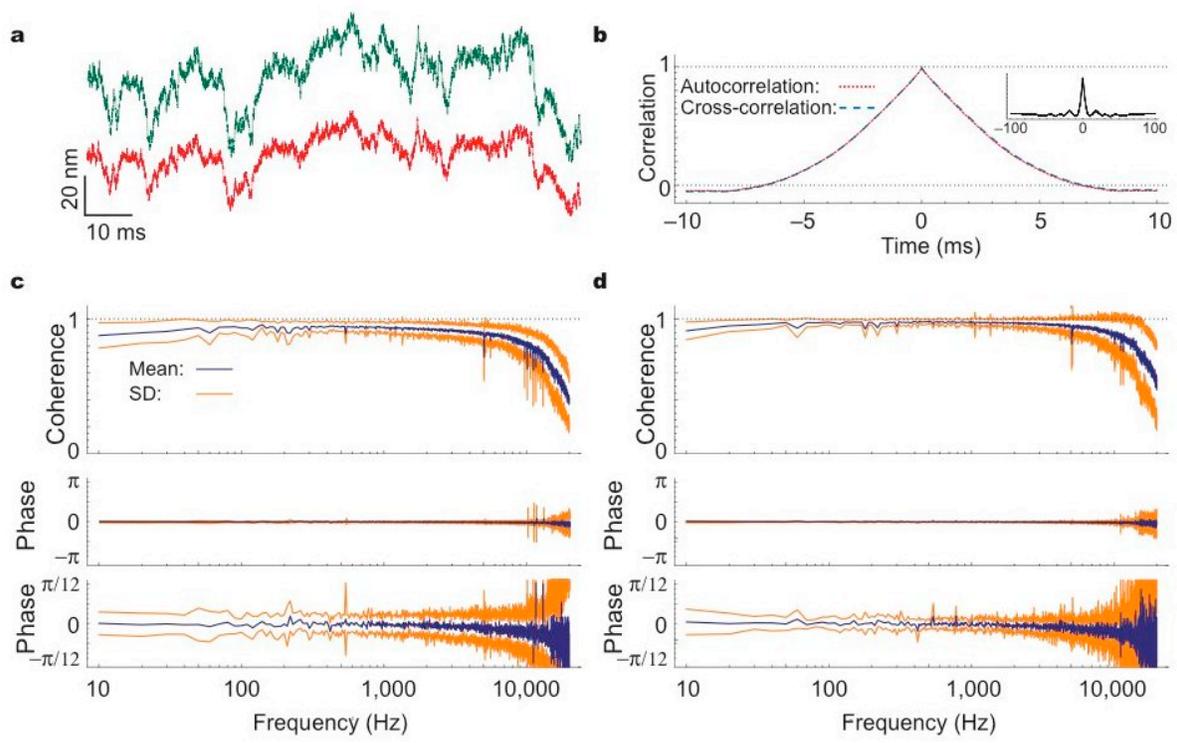

Figure 4

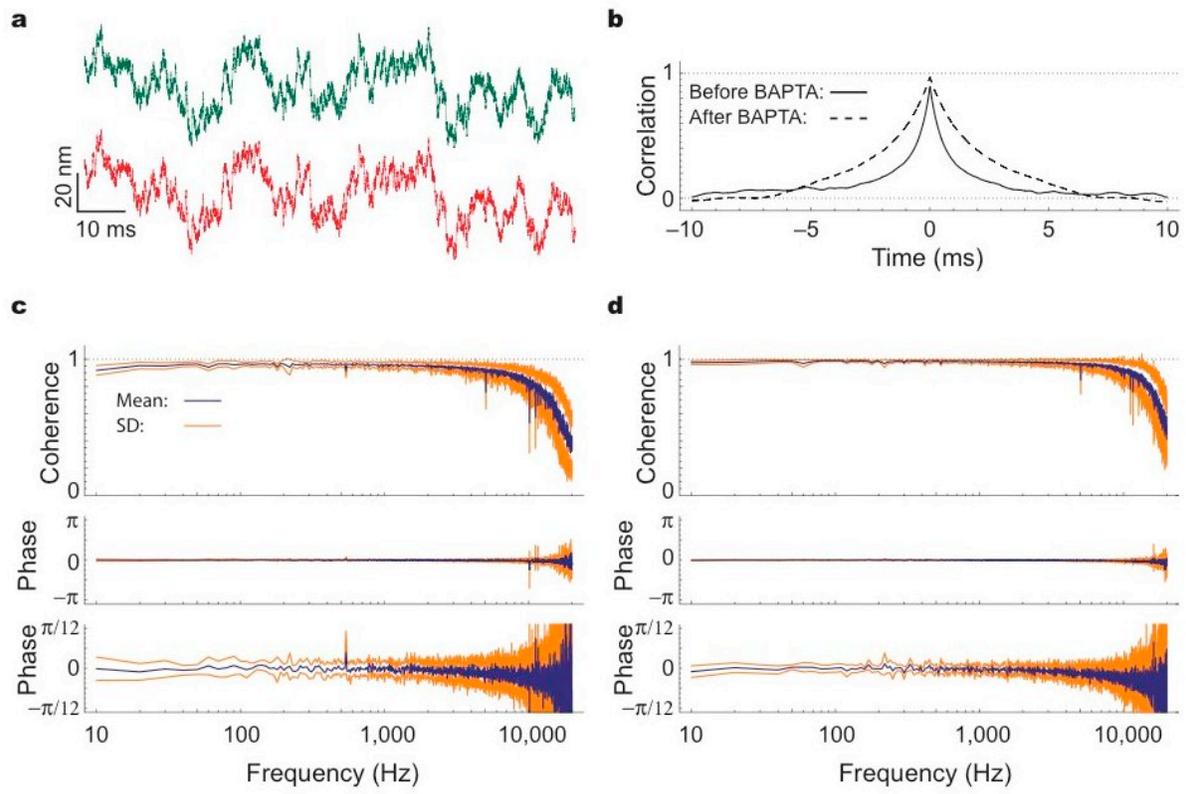

Figure 5

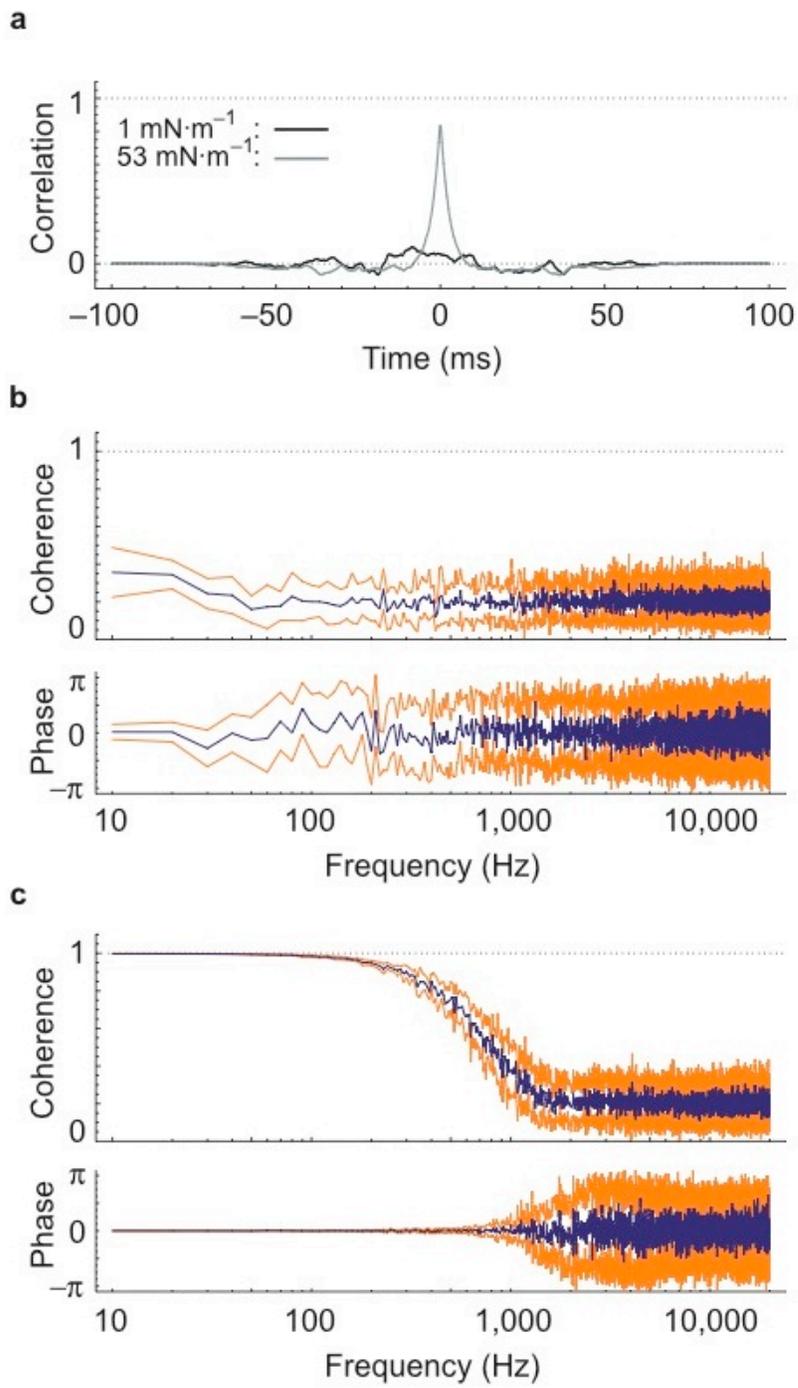

**a**

1 mN·m⁻¹ :
53 mN·m⁻¹ :

Figure 6